\documentclass[12pt]{article}
\usepackage{epsf}
\usepackage{amsmath}

\usepackage{graphics}
\usepackage{cite}

\begin{titlepage}

\begin{document}

\hfill August 2024

\begin{center}

{\bf \LARGE Avoidance of Schwinger Mechanism in Electromagnetically Accelerating Universe}\\
\vspace{2.5cm}
{\bf Paul H. Frampton}\footnote{paul.h.frampton@gmail.com}\\
\vspace{0.5cm}
{\it Dipartimento di Matematica e Fisica "Ennio De Giorgi",\\ 
Universit\`{a} del Salento and INFN-Lecce,\\ Via Arnesano, 73100 Lecce, Italy.
}

\vspace{1.0in}
\end{center}

\begin{abstract}
\noindent
\end{abstract}
\noindent
In quantum electrodynamics, Schwinger showed in 1951
that in a uniform electric field the vacuum may
spontaneously produce electron-positron pairs.
For this to occur, the required electric field is exceptionally large, $\sim 10^{18}$ N/C.
In the Electromagnetically Accelerating Universe (EAU)
model, such an electric field would be undesirable because it discharges 
the Primoridal Extemely Massive
Naked Singularities (PEMNSs) whose mutual  repulsion
causes the acceleration. This
leads to a constraints
on the
charges and masses  of the PEMNSs which must be satisfied in an EAU model.

\end{titlepage}

\noindent
\section{Introduction} 
\bigskip

\noindent
In the present paper, we address the issues of the mass
and charge of the Primordial Extremely Massive Naked
Singularities (PEMNSs) hypothesised in
using the novel Electromagnetically Accelerating Universe (EAU)
model.

\bigskip

\noindent
In the EAU model \cite{FramptonPLB,FramptonMPLA} 
the main idea is that electromagnetism
dominates over gravitation in the explanation of the accelerating
cosmological expansion. This idea takes us beyond the first paper\cite{Einstein}
which applied general relativity to theoretical cosmology.
This is not surprising, since in 1917 that author was obviously unaware
of the fact \cite{Perlmutter,Riess} discovered only in 1998 that the rate of cosmological expansion
is accelerating.

\bigskip

\noindent
We study in turn the avoidance of the Schwinger mechanism, the condition
for super extremality and dominance of Coulomb repulsion.

\bigskip
\bigskip

\section{The Model}

\noindent
The EAU model is fully relativistic and employs the classical
theories of electrodynamics and of general relativity. A key
ingredient is the assumption that there exist in the universe
a significant number of extremely massive objects with masses
in excess of a trillion solar masses which carry unscreened
like-sign electric charges. They are denoted by the acronym
PEMNSs standing for Primordial Extremely Massive
Naked Singularities.

\bigskip

\noindent
To be specific, we shall initially study an example of the EAU model
where the assumed common mass of the PEMNSs is $10^{12} M_{\odot}$
and their common negative electric charge is $-10^{32}$ Coulombs.

\bigskip

\noindent
To clarify the EAU model, it is worth stepping back and looking at the
big picture concerning the fundamental forces and comparing the
length scales in the conventional $\Lambda CDM$ model with those
in the present EAU model. In both models the short-range strong
interactions dominate the physics inside the atomic nucleus at
scales up to $10^{-15}$ metre. The nuclear weak force operates
at even smaller scales. In both models, when we discuss the larger scales
associated with atoms, molecules and all of chemistry it
is electromagnetic forces which dominate. At all these microscopic
length scales, gravitation can safely be neglected as infinitesimal

\bigskip

\noindent
At macroscopic scales, gravitation becomes dominant at everyday
scales and this continues at the scales characterising stars and galaxies.
In the $\Lambda CDM$ model dominance by gravity continues
to the largest cosmological scales. By contrast, in the EAU model,
at a scale larger than galaxies, at 100 Mpc and above, 
electromagnetism instead controls the dynamics. This is a key
difference between the EAU and $\Lambda CDM$ models.

\bigskip

\noindent
The PEMNSs are described in general relativity by a Reissner Nordstrom (RN)
metric which accommodates mass and charge. In order that the
Coulomb repulsion exceeds the gravitational attraction, the RN
metric must be super extremal and hence has a naked singularity.
This contradicts a cosmic censorship conjecture made in 1969
\cite{Penrose} but such a conjecture was unproven.

\bigskip

\noindent
In the $\Lambda CDM$ model, the accelerated cosmic expansion
discovered \cite{Perlmutter,Riess} in 1998 is caused by a mysterious
dark energy which somehow creates a counterintuitive repulsive
gravitational force. In the EAU model, it is caused by electrical
repulsion between the charged PEMNSs.

\bigskip

\noindent
Whereas the cosmological constant stays at its initial value
in $\Lambda CDM$, in the EAU model it is time dependent,
as we shall discuss later in this article.

\section{Avoidance of Schwinger Mechanism}

In 1951, Schwinger \cite{Schwinger1951} showed in the context of QED that
in a strong enough constant electric field, electron
positron pairs are created from the vacuum. This could lead to the
discharge of the PEMNSs by attracting the positron.

\bigskip

\noindent
According to \cite{Schwinger1951}, to avoid the Schwinger
mechanism requires the the electric field $|E|$ satisfies

\begin{equation}
|E| < 10^{18}  N/C
\label{Schwinger}
\end{equation}

\noindent
Let is consider a PEMNS with mass $10^p M_{\odot}$ and
electric charge $|Q| = 10^q$  Coulombs. A characteristic
scale is the Schwarzschild radius $R_S \sim 3 \times 10^{p+3}$ metres
at which the electric field is
\begin{equation}
|E| = \frac{k_e |Q|}{R_S^2} \sim 10^{q-2p+3}  N/C
\label{Field}
\end{equation}
using the electric constant $k_e = 9\times 10^9 n m^2 C^{-2}$.

\bigskip

\noindent
Consistency of Eq.(\ref{Field}) with Eq.(\ref{Schwinger}) requires that
\begin{equation}
(q-2p) < 15
\label{consistency}
\end{equation}

\noindent
Using the value $p=12$ and $q=32$ used in the EAU model
\cite{Frampton2024} we see that the electric field associated
with the Schwinger mechanism is avoided by several orders
of magnitude.

\section{Condition for Super Extremality}

In the Reissner-Nordstrom metric \cite{Reissner} describing an electrically
charged black hole, there are two independent lengths
depending respectively on the mass and the charge.

\begin{equation}
r_S = \frac{2GM}{c^2}
\label{rS}
\end{equation}

\noindent
and

\begin{equation}
r_Q^2 = \frac{k_e Q^2 G}{c^4}
\label{rQ}
\end{equation}

\bigskip
\noindent
Extremality depends on the relative sizes of $r_S$ and $(2r_Q)$.
If $r_S > 2r_Q$, the solution is sub extremal with two horizons.
When $r_S = 2 r_Q$, it is extremal with one unique horizon
at the Schwarzschild radius, $r_Q$.

\bigskip

\noindent
Super extremality occurs
when $2 r_q > r_S$. From Eqs. (\ref{rS}) and (\ref{rQ}) this requires
that
\begin{equation}
\sqrt{k_e} Q \geq \sqrt{G} M
\label{superex}
\end{equation}
which, for mass $M=10^p M_{\odot}$ and $Q=10^q$, gives
\begin{equation}
(q-p) \geq 20
\label{superex2}
\end{equation}

\noindent
With Eq. (\ref{superex2}) there is no horizon and hence a naked singularity.

\bigskip
\bigskip

\section{Dominance of Coulomb Repulsion}

Let us consider two identical PEMNSs each with mass
$M = 10^p M_{\odot}$ and electric charge $Q = - 10^q$ Coulombs, separated
by distance $R$. Then the Coulomb repulsion will dominate provided that
\begin{equation}
\frac{k_e Q^2}{R^2}  \geq \frac{G M^2}{R^2}
\label{dominance}
\end{equation}

\bigskip

\noindent
Using the values $k_e = 9 \times 10^9 N m^2 C^{-2}$,
$G = 6.67 \times 10^{-11} m^3 kg^{-1} s^{-2}$,
$M_{\odot} = 2 \times 10^{30}kg$ 
Eq.(\ref{dominance}) translates into
\begin{equation}
(q - p) \geq 20
\label{Coulomb}
\end{equation}

\bigskip

\noindent
For the EAU model \cite{Frampton2024}, Eq.(\ref{Coulomb})
is satisfied.  As the PEMNQ mass is increased beyond $10^{12} M_{\odot}$,
we learn from Eq.(\ref{Coulomb}) that the electric charge
must increase at least linearly in the mass.

\section{Time dependence of the Cosmological \\''Constant"}

\bigskip

\noindent
In the EAU model, the dark energy which has constant
time-independent density in $\Lambda CDM$ has been replaced
by electrically charged PEMNSs which is a part of dark matter.
Because of this, the cosmological term in the Friedmann
equation becomes time dependent.

\bigskip

\noindent
The particle masses, coupling constants and
mixing angles in the standard model of particle theory
are widely assumed to all be constants enshrined in
fundamental laws of physics. The cosmological
constant need not be of this immutable type and
it is quite plausible for it to be time dependent.

\bigskip

\noindent
Let us discuss the change in $\Lambda (t)$ between
$t= t_{DE} =9.8Gy$ when the accelerated cosmic expansion
began, until the present time $t = t_0 = 13.8 Gy$. In the
EAU model
\begin{equation}
\Lambda(t) \propto \left( \frac{1}{a(t)} \right)^3
\label{lambda}
\end{equation}
and the scale factor, normalised to $a(t_0)=1$, is given by
\begin{equation}
a(t) = \exp [ H_0 (t - t_0) ]
\label{scale}
\end{equation}
in which $H_0^{-1} =13.8 Gy$.

\bigskip

\noindent
From Eq.(\ref{lambda},\ref{scale}), it follows that
\begin{equation}
\left( \frac{\Lambda(t_0)}{\Lambda(t_{DE})} \right) = 0.36
\label{ratio}
\end{equation}

\bigskip

\noindent
This significant reduction in the cosmological constant could be
investigated by observations which can reconstruct the expansion
history of the universe.

\section{Summary}

\noindent
The EAU model is sufficiently revolutionary that it is easy to
suspect that there must be a fatal flaw. Therefore it is well worth
discussing at length the physics resulting from our
three previous short sections.

\bigskip

\noindent
In the EAU model, since a central r\^{o}le is played by the PEMNSs,
it is important to attribute to them masses $M = 10^p M_{\odot}$
and electric charges $10^q $ Coulombs. The values of $p$ and
$q$ exponents are our principal topic here. Values were chosen
in 2022 \cite{FramptonPLB} and 2023 \cite{FramptonMPLA}.

\bigskip

\noindent
We assume that the PEMNSs have a mass at least equal to
the Milky Way, including it dark matter halo, {\it i.e.} a mass
of one trillion $10^{12}$ Suns. Because they are black hofes, or better     
naked singularities, they are far more concentrated and have
a physical size of about one light year, compare to the hundred
thousand light years of the Milky Way.

\bigskip

\noindent
If the Schwinger mechanism had been operative in the vicinity
of the PEMNSs, it would provide a fatal flaw of the EAU model
because pair production in the vacuum would enable the
PEMNSs to lose their electric charge and thus fail to drive
the accelerating expansion. But the mechanism requires a
large electric field, too large to be generated by the PEMNSs
contained in the EAU model\cite{Frampton2024} which used
the parameters $p=12$ and $q=32$.

\bigskip

\noindent
If we increase the mass, keeping to a linear increase in charge,
for exmple to $p=23$ and $q=43$, the pair creation is suppressed
according to Eq.(\ref{consistency}) by several further orders of
magnitude. Thus the biggest risk of the triggering the Schwinger mechanism
comes with the minimal value $p=12$ and even there is avoided by
an electric field at least five orders of magnitude too small.

\bigskip

\noindent
In summary, the Schwinger mechanism does not provide a fatal
flaw for the EAU model.

\bigskip

\noindent
The subsequent two sections about respectively the condition
for super extremality and the dominance of Coulomb repulsion
lead to similar, actually identical, calculations. This is in itself
very interesting.

\bigskip

\noindent
This is one of the motivations for naked singularities, since
we learn that two RN extremal black holes exert no force on
one another yet when they increase their charge to become
super extremal, assuming all the PEMNS s have the same
sign electric charge, the Coulomb repulsion dominates over
the gravitational attraction.

\bigskip

\noindent
If we increase the PEMNS mass, it leads to two changes., .
Firstly, the total entropy increases. As we increase $p$
from $p=12$ to, say, $p=23$ the number of PEMNSs,
assuming the total mass is $10^{23} M_{\odot}$, the number
decreases from $n=10^{11}$ to $n=1$. The total entropy
increases from $10^{113}$ to $10^{124}$ eventually
saturating the maximal holographic bound.

\bigskip

\noindent
Meanwhile, by making these changes, the charge asymmetry
$\epsilon_Q$ remains unaltered assuming that, as suggested
earlier, $q$ is increased linearly with $p$. The incremental
increase to the total charge is $\Delta Q = 10^{43}$ Coulombs
compared to   the total charge of electrons, or protons, in the visible universe
$Q \sim 10^{61}$ Coulombs and hence $\epsilon_Q \sim 10^{-18}$.

\bigskip

\noindent
There are at least two possibilities regarding the charge asymmetry.
It could be an initial condition at the Big Bang or, if not, we must
live with $\epsilon_Q \neq 0$ at present.

\bigskip

\section{How to test the EAU model}

\bigskip

\noindent
Ideally, one would like to find evidence for the existence of 
PEMNSs. One valiant attempt to do this was presented in
\cite{Wagner} but it is unclear whether it is practicable.

\bigskip

\noindent
To support the general approach, it would be encouraging
if the PIMBHs in the Milky Way could be found using the Rubin
Observatory by microlensing of stars in the Magellanic Clouds.


\newpage





\begin{thebibliography}{ccc}

\bibitem{FramptonPLB}
P.H. Frampton,\\
{\it Electromagnetic Accelerating Universe}.\\
Phys. Lett. {\bf B835,} 137480 (2022).\\
{\tt arXiv:2210.10632[physics.gen-ph]}.

\bibitem{FramptonMPLA}
P.H. Frampton,\\
{\it  A Model of Dark Matter and Energy}.\\
Mod. Phys. Lett. {\bf A38,}  2350032 (2023).\\
{\tt arXiv:2301.10719[physics.gen-ph]}.

\bibitem{Einstein}
A. Einstein,\\
{\it Cosmological Considerations in the \\
General Theory of Relativity}.\\
Sitz.Preuss.Akad. {\bf 1917,} 142 (1917).

\bibitem{Perlmutter}
S. Perlmutter, {\it et al.} (Supernova Cosmology Project Collaboration)\\
Astrophys.J. {\bf 517,} 565 (1999).\\
 {\tt arXiv:astro-ph/9812133 [astro-ph]}.
 
\bibitem{Riess}
A.G. Riess, {\it et al.} (Supernova Search Team),\\
Astron.J. {\bf 116,} 1009 (1998).\\
{\tt  arXiv:astro-ph/9805201 [astro-ph]}.

\bibitem{Penrose}
R. Penrose,\\
Riv. Nuovo Cim. {\bf 1,} 252 (1969).

\bibitem{Schwinger1951}
J.S. Schwinger, \\
{\it On Gauge Invariance and Vacuum Polarization.}\\
Phys. Rev. {\bf 82,} 664 (1951).

\bibitem{Reissner}
H. Reissner,\\
{\it \"{U}ber die Eigengravitation des Elektrischen Feldes \\
nach der Einsteinschen Theorie}.\\
Annalen Physik {\bf 355,} 106 (1916).\\
G. Nordstrom,\\
{\it On the Energy of the Gravitational Field \\
in Einstein's Theory}.

\bibitem{Frampton2024}
P.H. Frampton,\\
{\it Status of Electromagnetic Accelerating Universe}.\\
In {\it Particle Theory and Theoretical Cosmology}, Festschrift \\
for 80th anniversary
of P.H. Frampton. To appear in \\
{\it Entropy} journal, and Special Edition, MDPI Books (2004).\\
{\tt arXiv:2404.08999[gr-qc]}
 \newpage
\bibitem{Wagner}
J. Wagner,\\
{\it Observables for Moving, Stupendously Charged and \\Massive
Primordial Black Holes}.\\
{\tt arXiv:2301.12210[astro-ph.CO]}\\
J.Wagner,\\
{\it Observables for Super-Extremal Black Holes Challenging \\Cosmic Censorship
to Comprehend the Cosmological Constant}.\\
{\tt arXiv:2405.08057[gr-qc]}.


\end{thebibliography}
\end{document}